\documentclass[runningheads]{llncs}
\usepackage[T1]{fontenc}
\usepackage{graphicx}
\usepackage{textcomp}
\usepackage{amsmath,amsfonts}
\usepackage{algpseudocode}
\usepackage{mathtools}
\usepackage{array}
\usepackage{graphicx}
\usepackage{version}
\usepackage{comment}
\usepackage{color}
\usepackage{soul}
\usepackage[normalem]{ulem}
\usepackage{subcaption}
\usepackage{rotating}

\includeversion{paper}
\excludeversion{extended}
\newcommand{\RoleActions}{\mathit{RoleActions}}
\newcommand{\RoleScript}{\mathit{RoleScript}}
\newcommand{\START}{\mathit{Start}}
\newcommand{\snd}{\mathit{Snd}}
\newcommand{\rcv}{\mathit{Rcv}}
\newcommand{\ck}{\mathit{Check}}

\newcommand{\Ag}{\mathit{Ag}}

\newcommand{\surgicalaction}{\mathit{s\_ action}}

\newcommand{\LS}{\mathit{S}}
\newcommand{\AS}{\mathit{A}}
\newcommand{\N}{\mathit{N}}
\newcommand{\C}{\mathit{O}}
\newcommand{\clipsrequested}{\mathit{clips\_ requested}}
\newcommand{\notclipsrequested}{\mathit{not\_ clips\_ requested}}
\newcommand{\cutdone}{\mathit{cut\_ done}}
\newcommand{\notcutdone}{\mathit{not\_cut\_ done}}
\newcommand{\cut}{\mathit{cut}}
\newcommand{\clipsapplied}{\mathit{clips\_ applied}}
\newcommand{\notclipsapplied}{not\_ clips\_ applied}
\newcommand{\apply}{\mathit{apply}}
\newcommand{\request}{\mathit{request}}
\newcommand{\clips}{\mathit{clips}}
\newcommand{\clipsprovided}{\mathit{clips\_ provided}}
\newcommand{\notclipsprovided}{\mathit{not\_ clips\_ provided}}
\newcommand{\provide}{\mathit{provide}}
\newcommand{\ureter}{\mathit{ureter}}
\newcommand{\pull}{\mathit{pull}}
\newcommand{\VDandSV}{\mathit{VD\_ and\_ SV}}
\newcommand{\dissect}{\mathit{dissect}}
\newcommand{\pedicle}{\mathit{pedicle}}
\newcommand{\cauterise}{\mathit{cauterise}}
\newcommand{\lookfor}{\mathit{look\_ for}}
\newcommand{\enter}{\mathit{enter}}
\newcommand{\incise}{\mathit{incise}}
\newcommand{\coagulate}{\mathit{coagulate}}
\newcommand{\visceralfascia}{\mathit{visceral\_ fascia}}

\newcommand{\PFS}{\mathit{PFS}}

\newcommand{\NVB}{\mathit{NVB}}

\newcommand{\inspect}{\mathit{inspect}}
\newcommand{\DF}{\mathit{DF}}
\newcommand{\DFincised}{\mathit{DF\_ incised}}
\newcommand{\visceralfasciaincised}{\mathit{visceral\_ fascia\_ incised}}

\newcommand{\pedicledissected}{\mathit{pedicle\_ dissected}}
\newcommand{\pediclecauterised}{\mathit{pedicle\_ cauterised}}
\newcommand{\pediclefound}{\mathit{pedicle\_ found}}
\newcommand{\NVBpreserved}{\mathit{NVB\_ preserved}}
\newcommand{\capsulararteries}{\mathit{capsular \; arteries}}
\newcommand{\smallarteries}{\mathit{small \; arteries}}
\newcommand{\PFSentered}{\mathit{PFS\_ entered}}
\newcommand{\VDandSVpulled}{\mathit{VD\_ and\_ SV\_ pulled}}
\newcommand{\smallarteriescoagulated}{\mathit{small\_ arteries\_ coagulated}}
\newcommand{\NVBandSAfound}{\mathit{NVB\_ and\_ small\_ arteries\_ found}}

\newcommand{\stkout}[1]{\ifmmode\text{\sout{\ensuremath{#1}}}\else\sout{#1}\fi}
\newcommand{\secref}[1]{\S\ref{#1}}
\newcommand{\QMX}{\mathit{?X}}

\newcommand{\prem}{\mathit{prem}}
\newcommand{\conc}{\mathit{conc}}

\newcommand{\K}[2]{\mathit{K_{#1}^{#2}}}

\newcommand{\ST}[2]{\mathit{S_{#1}^{#2}}}

\newcommand{\NewTransitionsShort}[5]{
\begin{align*}
    \mathit{#1} & \xrightarrow{#2,} \\
    & \qquad \xrightarrow{#3,} \\
    & \qquad\qquad \xrightarrow{#4} 
    #5 
\end{align*}
}

\newcommand{\NewTransitionsShortend}[4]{
\begin{align*}
    \mathit{#1} & \xrightarrow{#2} \\
    & \qquad \xrightarrow{#3} #4
\end{align*}
}

\newcommand{\NewTransitions}[6]{
\begin{align*}\footnotesize
    \mathit{#1} & \xrightarrow{#2,} \\
    & \qquad \xrightarrow{#3,} \\
    & \qquad\qquad \xrightarrow{#4,} \\
    & \qquad\qquad\qquad \xrightarrow{#5}  
    #6 
\end{align*}
}

\usepackage{etoolbox}

\AtBeginEnvironment{property}{\footnotesize}
\AtEndEnvironment{property}{}

\begin{document}

\title{A Formal Approach For Modelling And Analysing Surgical Procedures \\ 
(Extended Version)\thanks{This is an extended version of the paper ``A Formal Approach For Modelling And Analysing Surgical Procedures'' that appeared in the proceedings of The 20th International Workshop on Security and Trust Management (STM 2024). This research was funded by the UK Research and Innovation Trustworthy Autonomous Systems Hub (EP/V026801/2).}}

\titlerunning{A Formal Approach For Modelling And Analysing Surgical Procedures}

\author{Ioana Sandu\inst{1} \and
Rita Borgo\inst{1} \and 
Prokar Dasgupta\inst{2} \and \\
Ramesh Thurairaja\inst{3}
\and
Luca Vigan\`o\inst{1}
}
\author{Ioana Sandu\inst{1}\orcidID{0009-0005-4512-7325} \and
Rita Borgo\inst{1}\orcidID{0000-0003-2875-6793} \and \\
Prokar Dasgupta\inst{2}\orcidID{0000-0001-8690-0445} \and
Ramesh Thurairaja\inst{3}
\and \\
Luca Vigan\`o\inst{1}\orcidID{0000-0001-9916-271X}
}

\authorrunning{I.~Sandu et al.}
\institute{Department of Informatics, King's College London, UK \and
Peter Gorer Department of Immunobiology, King's College London, UK 
\and Urology Department, Guy's \& St Thomas' Hospital NHS Foundation Trust, UK \\
\email{\{ioana.sandu, rita.borgo, prokar.dasgupta, luca.vigano\}@kcl.ac.uk}, \email{Ramesh.Thurairaja@gstt.nhs.uk}
}

\maketitle              

\begin{abstract}
Surgical procedures are often not ``standardised'' (i.e., defined in a unique and unambiguous way), but rather exist as implicit knowledge in the minds of the surgeon and the surgical team. This reliance extends to pre-surgery planning and effective communication during the procedure. We introduce a novel approach for the formal and automated analysis of surgical procedures, which we model as security ceremonies, leveraging well-established techniques developed for the  analysis of such ceremonies. 
Mutations of a procedure are used to model variants and mistakes that members of the surgical team might make. Our approach allows us to automatically identify violations of the intended properties of a surgical procedure. 

\end{abstract}


\section{Introduction}
\label{sec:introduction}

\emph{Context and Motivation.} 
This paper is the result of a collaboration between computer scientists and clinician scientists. The collaboration commenced with the live observation of a robot-assisted prostatectomy and cystectomy, leading to in-depth discussions on the actual execution of surgical procedures.
These emphasised that much of a surgical procedure is often in the heads of the surgeon and of the members of the surgical team. This reliance on internalised knowledge hinges on two critical activities: (1)~comprehensive pre-surgery discussions between the team members, (2)~effective communication throughout the procedure. 
Recognising the potential for errors in both activities, which could jeopardise patient safety, 
\emph{surgical process models (SPMs)} have been proposed to represent surgical procedures. These models
offer ``simplified, formal, or semi-formal representations of a network of surgery-related activities''~\cite{neumuth2017surgical}.\footnote{See~\cite{Guillonneau2000,Martini2020,Villers2017,zhou2020transvesical} for SPMs for robot-assisted prostatectomies and similar procedures.}
SPMs draw upon concepts from workflow management and computer science, and (often) provide a representation of a surgical procedure that can be communicated to team members as well as to other surgeons so they may follow the same steps. However, even in the case of more formal SPMs, little to no attention has been devoted to using SPMs to reason about procedures, particularly in the context of (1) and (2).

\emph{Contributions.} 
We propose a different approach but one still anchored in computer science and, more specifically, cybersecurity: we formally model and reason about surgical procedures by representing them as security ceremonies.

Modelling a surgical procedure as a security ceremony brings some important advantages. It provides conceptual clarity and allows one to represent the procedure as a message sequence chart that can be published and shared with others. It also provides a structured framework for reasoning as it enables us to adapt to surgical procedures established methodologies and automated approaches developed for the formal analysis of security ceremonies and their properties. 

A \emph{security protocol}, sometimes also called \emph{cryptographic protocol}, is essentially a communication protocol (an agreed sequence of actions performed by two or more communicating agents in order to accomplish some mutually desirable goals) that makes use of cryptographic techniques, allowing the communicating agents to satisfy one or more security properties (such as authentication, confidentiality of data or integrity of data).
A \emph{security ceremony} expands a security protocol to include human nodes alongside computer nodes, with communication links that comprise user interfaces, human-to-computer and human-to-human communication, and transfers of physical objects that carry data~\cite{Vigano22}. Hence, a ceremony's analysis should include, in particular, the mistakes that human agents might make when they communicate with each other or when they execute their tasks.

Modelling a surgical procedure as a security ceremony thus allows us to consider \emph{mutations} of the ceremony/procedure that formalise possible mistakes made by members of the surgical team. For instance, we formally model that the actions of the surgical team should not cause internal bleeding or endanger the patient, and our approach allows us to capture violations of this property, e.g., situations where a surgeon performs an internal incision without the assistant applying clips to prevent bleeding.\footnote{We focus here on mistakes carried out by the human agents, but mistakes caused by robotic agents could also be considered similarly.} To automatically identify that such mistakes violate the intended properties of the procedure, we adapt and extend the mutation-based analysis approach proposed in~\cite{SempreboniVigano2022} (although we use a different tool, namely UPPAAL, \url{https://uppaal.org}, as it provides greater visual simplicity).
Furthermore, mutations provide the means for both researchers and surgeons to explore variants of the procedure (e.g., alterations in the order of actions) and check whether they lead to property violations without having to perform the variant on a live patient. 

To illustrate our approach, we consider two different stages of a laparoscopic prostatectomy procedure that is described informally in~\cite{Guillonneau2000}: the cutting stage and the lateral dissection stage, in which the lateral surfaces of the prostate are dissected. We use the cutting stage as a running example that allows us to introduce our formal modelling approach and detail the mutations and matched mutations, whereas the lateral dissection stage provides a more intricate sequence of actions and shows how our approach can go beyond the simple example of performing a cut.
We chose to consider \cite{Guillonneau2000} as this paper provides one of the most comprehensive and detailed descriptions of the different stages of a laparoscopic prostatectomy. However, the description in~\cite{Guillonneau2000}, as is standard in such papers, is completely textual (with only some anatomical illustrations) and thus informal and prone to misunderstandings, so providing a formal model as we do is already a valuable contribution, followed by the contribution provided by the formal analysis. Note that \cite{Guillonneau2000} is now more than 20 years old and some specifics of the standard laparoscopic prostatectomy procedure might have changed in the meantime (cf.~some of the newer papers that we discuss in the next section) but adapting our models and analysis accordingly would be quite straightforward.

\emph{Structure.} 
In \secref{sec:background}, we discuss background and related work. 
We present our formal model in \secref{sec:model}, the mutations in \secref{sec:mutations}, and the formal analysis in \secref{sec:formal-analysis}. We draw conclusions in \secref{sec:conclusions}. Additional details are provided in appendix.

\section{Background and Related Work}
\label{sec:background}

\emph{Robotic-assisted surgery (RAS)} has transformed the conventional operating room by introducing changes that include increased spatial requirements due to equipment and the physical separation of console surgeons from patients and team. 
In contrast to traditional arrangements~\cite{sexton2018anticipation}, the configuration of RAS may hinder interpersonal cues and lead to potential miscommunication. 

Our approach proposes to reason about surgical procedures by conceptualising them as security ceremonies, which offer an explicit representation of human agents and their communications with other agents (human or not)~\cite{Curzon2017,SempreboniVigano2022}. This perspective enables us to systematically incorporate and reason about human mistakes in the context of RAS (in fact, surgical procedures of any kind), but we could similarly model robot agents in RAS and other features of such procedures.

This is important as the work system inherent in RAS encompasses various elements such as the patient, surgery type, surgical goals, tasks contributing to those goals, patient-related factors, and situational factors. The integration of new technologies into the operating room has the potential to significantly alter the prerequisites for effective teamwork, procedural workflows, and individual skills~\cite{Souders2019-tz}. The distinctive setup of RAS introduces new challenges in maintaining situational awareness, team coordination, and information exchange~\cite{weber2018effects}. 
Hence, effective communication is crucial for maintaining a surgeon's situational awareness, especially when operating from the console~\cite{weigl2018associations}. Communication, a historically recognised source of disruption in surgeries, undergoes fundamental changes in RAS due to the relocation of the surgeon from the operating table, and the impact of workflow disruptions/interruptions has been explored in, e.g., 
\cite{weber2018effects,weigl2018associations}. Specific verbal/non-verbal 
cues are crucial for team coordination~\cite{catchpole2019human}, and studies have delved into the influence of anticipation and teamwork in RAS~\cite{sexton2018anticipation}.  

Excellent methods for conventional laparoscopic radical prostatectomy have been described in, e.g.,
\cite{Guillonneau2000,Martini2020,zhou2020transvesical}, 
but there is currently no standardised surgical technique for robot-assisted radical prostatectomy. 

In this paper, we introduce our approach by initially demonstrating its efficacy with a simple stage of the procedure, the cutting stage, followed by a more intricate stage, the dissection of the lateral surfaces of the prostate.
The latter stage is pivotal because preserving the neurovascular bundles is paramount for ensuring a successful surgical outcome for patients who aim to maintain postoperative potency.
Failure to preserve these bundles could significantly impact such patients' recovery. Various approaches to nerve-sparing prostatectomies are discussed in \cite{Moschovas2022-ff}. {Denonvilliers' fascia} is a crucial structure covering the posterior surface of the prostate and separating it from the rectum. It plays a vital role in the confinement of cancer within the prostate and facilitating an operation without damaging the nerves responsible for erectile function and continence, while ensuring the removal of all neoplastic tissue \cite{Tzelves2022-gm}. Therefore, this stage not only demonstrates the close collaboration between the surgeon and the assistant but also allows us to reason about one of the key factors contributing to a successful outcome and recovery. 

In the following, we demonstrate how our approach allows us to reason about mistakes that team members might make, and briefly discuss how it allows us to reason about variants of a procedure (i.e, different ways of performing it).


\section{Formal Model}
\label{sec:model}

\begin{figure}[ht!]
\begin{minipage}[t]{0.45\textwidth}
\centering
\includegraphics[scale=1]{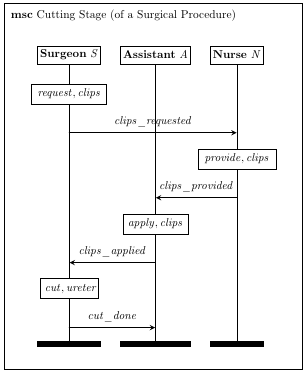}
\subcaption{MSC of the Cutting Stage}\label{fig:msc_a}
\end{minipage}
\begin{minipage}[t]{0.54\textwidth}
\centering\includegraphics[scale=0.12]{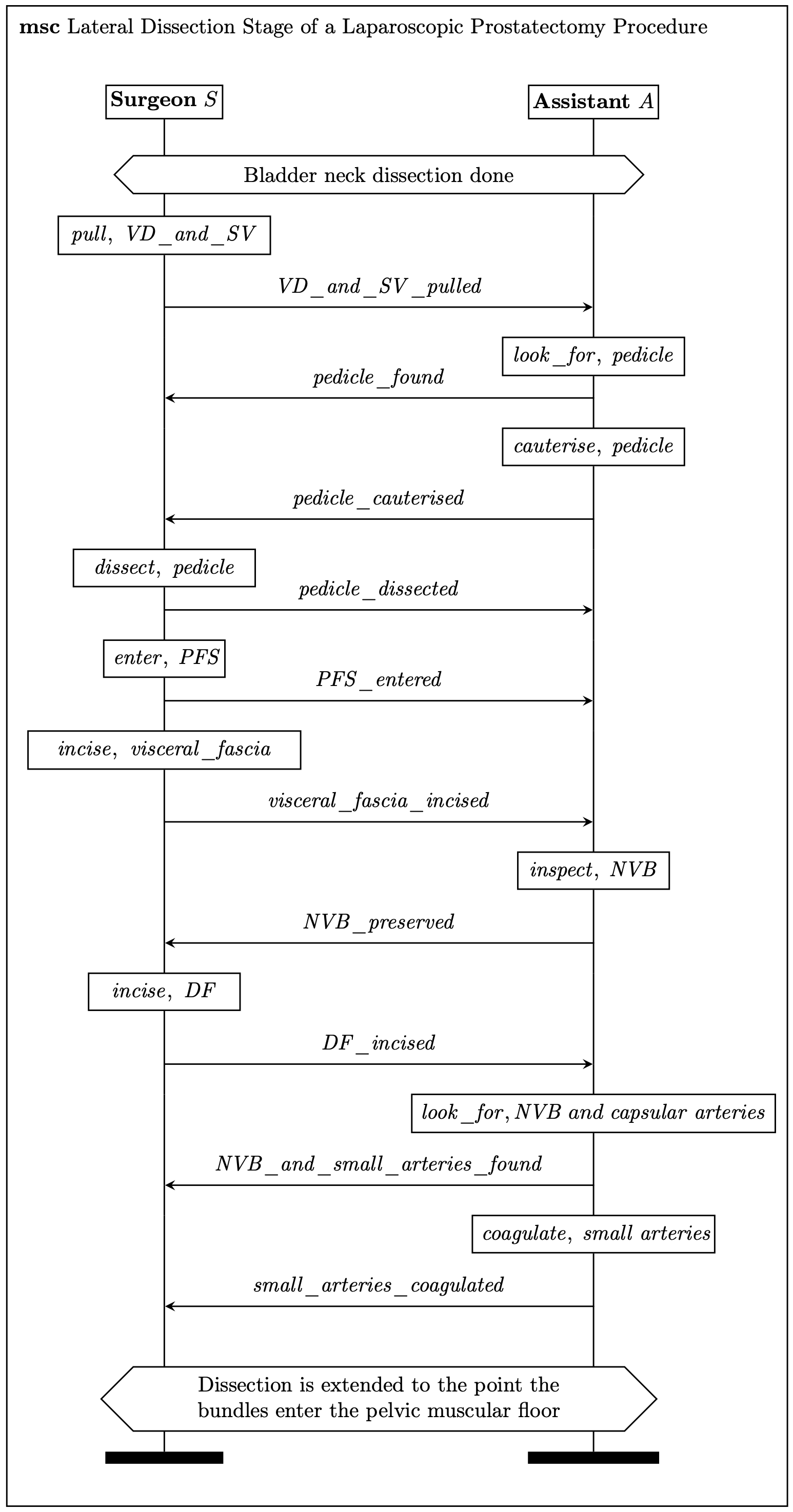}
\subcaption{MSC of the Lateral Dissection Stage}\label{fig:msc_b}
\end{minipage}
\caption{MSCs of Two Stages of a Laparoscopic Prostatectomy Procedure}\label{fig:msc}
\end{figure}

\label{sec:model-introduction}
In a surgical procedure, multiple agents collaborate through a series of orchestrated actions and message exchanges to execute their tasks seamlessly as a cohesive team. We thus model a surgical procedure as a sequence of actions and messages exchanged so that other actions can occur. As a concrete example, we provide a formal model (and security analysis) of two stages of a laparoscopic prostatectomy procedure that is described informally in~\cite{Guillonneau2000}.
The
\emph{message sequence chart (MSC)} in Fig.~\ref{fig:msc_a} shows the cutting stage of this procedure, where three agents, a surgeon $\LS$, an assistant $\AS$ and a nurse $\N$, collaborate to carry out an internal incision on a patient.\footnote{Note that this cutting stage is quite general and could be applied also to other surgical procedures and not just to a prostatectomy.} Fig.~\ref{fig:msc_b} shows the MSC that we have drawn for another stage of the  procedure, i.e, the dissection of the lateral surfaces of the prostate, which we abbreviate as \emph{lateral dissection stage}, and where $\mathit{VD}$, $\mathit{DF}$, $\mathit{SV}$ and $\mathit{NVB}$ abbreviate Denonvilliers' Fascia, vasa deferentes, seminal vesicles and neurovascular bundles, respectively.

Agents are often referred to also as \emph{roles}, and we will use ``agent'' and ``role'' interchangeably.
In an MSC, each agent is defined as a process characterised by a series of surgical actions (the boxes in each agent's vertical timeline) and messages they send (the horizontal arrows), confirming that they have carried out an action or requested other agents to carry out an action. 

\begin{extended}
\subsubsection{Messages}
\label{Subsec:messages}
\end{extended}
Agents can send and receive \emph{messages}. 
In adherence with the standard in formal modelling and analysis of security protocols and ceremonies (e.g., \cite{SempreboniVigano2022,Vigano22}), we define the \emph{algebra of messages} as $\mathcal{T}_\Sigma(\mathcal{V})$. The signature $\Sigma$ contains possibly disjoint sets of constants (e.g., representing agent names and other publicly known values) and $\mathcal{V}$ is a countably infinite set of variables.
$\Sigma$ can easily be extended to include function symbols to formalise symmetric and asymmetric decryption and other cryptographic operators. 

Given the set $M$ of all messages that can be built according to the algebra, we define for each agent $\Ag$ a set $M_\Ag^s$ of messages $\Ag$ can send and a set $M_\Ag^r$ of messages $\Ag$ can receive.
In this paper, we only consider messages that can be defined as constants, as that is what our case study requires. For instance, for 
both stages of the procedure, we define these sets as follows:
\begin{displaymath}\footnotesize
\begin{array}{ll}
M^s_\LS{} \ = & \{\clipsrequested{}, \cutdone{}, \VDandSVpulled{}, \pedicledissected{}, \\
&\PFSentered{}, \visceralfasciaincised{}, \DFincised{}\}\} \\
M^r_\LS{} \ = & \{\clipsapplied{}, \pediclefound{}, \pediclecauterised{}, \NVBpreserved{}, \\
&\NVBandSAfound{}, \smallarteriescoagulated{} \} \\
M^s_{\AS} \ = & \{\clipsapplied, \pediclefound{}, \pediclecauterised{}, \NVBpreserved{}, \\
& \NVBandSAfound{}, \smallarteriescoagulated{}\} \\
\end{array}
\end{displaymath}
\begin{displaymath}\footnotesize
\begin{array}{ll}
M^r_{\AS} \ = & \{\clipsprovided{}, \cutdone{}, \VDandSVpulled{}, \pedicledissected{}, \\
& \PFSentered{}, \visceralfasciaincised{}, \DFincised{}\} \\
M^s_N \ = & \{\clipsprovided\} \\
M^r_N \ = & \{\clipsrequested\} \\
M \ \ \; = & M^s_\LS{} \cup M^r_\LS{} \cup M^s_{\AS} \cup M^r_{\AS} \cup M^s_N \cup M^r_N
\end{array}
\end{displaymath}
Our approach is flexible and can accommodate more complex messages, e.g., that contain random numbers and are encrypted, as is typical in security ceremonies. 
Moreover, we consider only honest agents who behave according to what the surgical procedure expects,
but below we will extend this to consider mistakes by agents (and we could extend further to consider dishonest agents who can do anything they want as is standard in formal analysis of security ceremonies).

\begin{extended}
\subsubsection{Role Script}
\label{SubSec:RoleScript}
\end{extended}
In formal analysis of security ceremonies, and thus in our approach here, agents are formalised as processes that represent 
the vertical lines in an MSC and that are often called role-scripts. 
A \emph{role-script} is a sequence of events $e \in T_{\Sigma \cup \RoleActions}$, where $\RoleActions = \{\snd, \rcv, \surgicalaction, \START \}$ is a set of action names with their respective arity as defined below. For example, the role-scripts of the agents of the cutting stage are as follows: 

\vspace*{-0.4cm}
\begin{displaymath}\footnotesize
\hspace*{-0.5cm}
\begin{array}{ccc}
    \begin{array}{l}
    \RoleScript_{\LS} = \\
    \quad [\START(\LS, \; \K{\LS}{0}) \\
    \quad \phantom{[}\surgicalaction(\LS, \; \request, \; \clips) \\
    \quad \phantom{[}\snd(\LS, \; \N, \; \clipsrequested) \\
    \quad \phantom{[}\rcv(\AS, \; \LS, \; \clipsapplied) \\
    \quad \phantom{[}\surgicalaction(\LS, \; \cut, \; \ureter ) \\
    \quad \phantom{[}\snd(\LS, \; \AS, \; \cutdone)]
    \end{array}
& \hspace*{-0.15cm}
    \begin{array}{l}
    \RoleScript_{\AS} = \\
    \quad [\START(\AS, \; \K{\AS}{0}) \\
    \quad \phantom{[}\rcv(\N, \; \AS, \; \clipsprovided) \\
    \quad \phantom{[}\surgicalaction(\AS, \; \apply, \; \clips) \\
    \quad \phantom{[}\snd(\AS, \; \LS, \; \clipsapplied) \\
    \quad \phantom{[}\rcv(\LS, \; \AS, \; \cutdone)]
    \end{array}
& \hspace*{-0.15cm}
\begin{array}{l}
    \RoleScript_{\N} = \\
    \quad [\START(\N, \; \K{\N}{0}) \\
    \quad \phantom{[}\rcv(\LS, \; \N, \; \clipsrequested) \\
    \quad \phantom{[}\surgicalaction(\N, \; \provide, \; \clips) \\
    \quad \phantom{[}\snd(\N, \; \AS, \; \clipsprovided)]
    \end{array}
\end{array}
\end{displaymath}
$\START(\Ag, \; K_\Ag^0)$ is the first event of a role-script and it takes place once, where $K_\Ag^0$ is the \emph{initial knowledge} of agent $\Ag$ at the beginning of the process (typically, it contains the names of the other agents and the messages $\Ag$ will send). 
$\Ag$'s \emph{knowledge} increases monotonically as it receives messages. 
$\snd$ and $\rcv$ events are of the form $\snd(\Ag_s, \; \Ag_r, \; m)$ and $\rcv(\Ag_s, \; \Ag_r, \; m)$, where $\Ag_s$ is the sender of the message, $\Ag_r$ is the receiver and $m$ is the message that is being sent or received.
In our model, the messages have only one recipient to indicate who the information concerns; this is primarily a modeling choice for clarity, but it can be easily adjusted so that multiple agents receive the same message. Moreover, we focus on secure communication channels between agents, but again our approach is open and extends to various types of channels (e.g., authenticated or insecure ones, which can be attacked by a dishonest agent). 

Surgical action events represent the actions each agent performs to ensure the progression of the surgery and are defined as $\surgicalaction (\Ag, \; a_\Ag, \C)$, agent $\Ag$ performs action $a_\Ag$ on or with object $\C$. For the two stages we are considering, 
$a_{\LS} \in \{\request, \; \cut, \; \pull, \; \dissect, \; \enter, \; \incise \}$
$a_{\AS} \in \{\apply, \; \lookfor, \allowbreak \cauterise, \; \inspect, \; \coagulate \}$ and $a_N \in \{\provide\}$.\footnote{We could represent cuts and incisions by means of a single action but we prefer to consider two distinct actions $\cut$ and $\incise$ to distinguish between actions that might use different instruments (e.g., scissors or scalpels).} The objects are
$\C \ \ = \{\clips, \; \ureter, \; \VDandSV, \; \pedicle, \; \PFS, \; \visceralfascia, \; \NVB, \; \DF, \allowbreak \capsulararteries, \; \smallarteries\}$, which are significant as the same action may be performed multiple times for different objects (e.g., one can request clips or a scalpel), or an action could be performed with the same object but at different stages of the procedure (e.g., clips can be requested during cutting or suturing).

\begin{extended}
\subsubsection{Execution Model}
\label{SubSec:ExecutionModel}
\end{extended}

Our approach is based on an \emph{execution model} that is defined by a \emph{multi-set rewriting system} like in many security protocol/ceremony analysis approaches and tools.
A \emph{state} is a multiset of facts that model resources, including the information that agents know and exchange. 
Formally, the state $\ST{\Ag}{i}=\{i; \; \K{\Ag}{i}\}$ of agent $\Ag$ 
is characterised by the state number $i$ and the knowledge $\K{\Ag}{i} = \;\K{\Ag}{0} \cup \{\mathit{messages\ received\ by}\; \Ag\}$ that $\Ag$ possesses at $i$, and $\ST{}{i} = 
\{i; \; \{\K{\Ag}{i} \mid \Ag \ \text{is\ an\ agent}\}\}$ represents the state of all agents at that point in the execution. 

A \emph{trace} is a finite sequence of multisets of role-actions and is generated by the application of \emph{state transition rules} of the form 
\begin{displaymath}\footnotesize
	\prem \xrightarrow[]{\mathit{finite\ sequence\ of\ role-actions\ and\ internal\ checks}} \conc
\end{displaymath}
which is applicable when the current state matches the premise $\prem$ and the internal checks on the messages received are satisfied. These checks are typically not displayed in a role-script but only act as guards. The rule's application produces the  conclusion $\conc$ (a new state) and records the instantiations of role-actions in the trace. 
For instance, for the cutting stage,
\NewTransitions
{\{1; \; \; \K{\LS}{1}, \K{\AS}{1}, \K{\N}{1}\}}
{\rcv(\LS, \; \N, \; \QMX)}
{\ck(\QMX \; = \; \clipsrequested)}
{\surgicalaction(\N, \; \provide, \; \clips)}
{\snd(\N, \; \AS, \; \clipsprovided)}
{\{2; \; \; \K{\LS}{2}, \K{\AS}{2}, \K{\N}{2}\}}
represents a transition from state $\ST{}{1}$ to state $\ST{}{2}$, where we split the arrow for readability and where we are numbering the states from $0$ to $5$ assuming that the cutting stage is the initial stage of the procedure (if it is not the initial stage, then the states will be numbered differently but still consecutively). Agent $\N$ receives a message $\QMX$ from $\LS$, checks the contents of that message, performs a surgical action and sends a confirmation message to $\AS$.
We write that agent $\Ag$ receives $\QMX$ to allow $\Ag$ to check, via $\ck(\QMX = m)$, that this is indeed the message $m$ that $\Ag$ was expecting. This check will become useful later as it will enable us to consider mistakes that agents could make, such as changing the contents of the message or sending the wrong message.\footnote{In security ceremonies, there typically is a $?$ also in front of the name of the sender in a $\rcv$ event. This allows one to consider an attacker that is claiming to be the sender. In this paper, we avoid doing so given that we are not yet explicitly considering an attacker, but our approach would easily allow us to do so.}

Nurse $\N$ can carry out only the transition above. Surgeon $\LS$'s can carry out 
\NewTransitionsShort
{\{0; \; \K{\LS}{0}, \K{\AS}{0}, \K{\N}{0}\}}
{\START}
{\surgicalaction(\LS, \; \request, \; \clips)} 
{\snd(\LS, \; \N, \; \clipsrequested)} 
{\{1; \; \K{\LS}{1}, \K{\AS}{1}, \K{\N}{1}\}}
\vspace*{-1cm}
\NewTransitions
{\{3; \; \K{\LS}{3}, \K{\AS}{3}, \K{\N}{3}\}}
{\rcv(\AS, \; \LS, \; \QMX)}
{\ck (\QMX \; = \; \clipsapplied)}
{\surgicalaction(\LS, \; \cut, \; \ureter )}
{\snd(\LS, \; \AS, \; \cutdone)}
{\{4; \; \K{\LS}{4}, \K{\AS}{4}, \K{\N}{4}\}}
and assistant $\AS$ can carry out
\NewTransitions
{\{2; \; \K{\LS}{2}, \K{\AS}{2}, \K{\N}{2}\}}
{\rcv(\N, \; \AS, \; \mathit{\QMX})}
{\ck(\QMX \; = \; \clipsprovided)}
{\surgicalaction(\AS, \; \apply, \; \clips)}
{\snd(\AS, \; \LS, \; \clipsapplied)}
{\{3; \; \K{\LS}{3}, \K{\AS}{3}, \K{\N}{3}\}}
\vspace*{-1cm}
\NewTransitionsShortend
{\{4; \; \; \K{\LS}{4}, \K{\AS}{4}, \K{\N}{4}\}}
{\rcv(\LS, \; \AS, \; \QMX)} 
{\ck(\QMX \; = \; \cutdone)} 
{\{5; \; \; \K{\LS}{5}, \K{\AS}{5}, \K{\N}{5}\}}

The expected sequence of actions is then given by the rules applied in the same order as their state numbers, mirroring the MSC. It captures the unaltered process, when all agents execute their tasks precisely as expected. No errors are occurring, but rather every action unfolds in a seamless sequence, with agents patiently awaiting messages from their predecessors before proceeding.

The transition rules of the agents for the lateral dissection stage are similar. 
For example, the following rule represents the incision of the 
DF between states $\mathit{i}$ and $\mathit{i+1}$ (we use $i$ to indicate the corresponding states as they occur in the full procedure when all stages are considered):

\NewTransitions
{\{i; \; \K{\LS}{i}, \K{\AS}{i}, \K{\N}{i}\}}
{\rcv(\AS, \; \LS, \; \mathit{\QMX})}
{\ck(\QMX \; = \; \NVBpreserved)}
{\surgicalaction(\LS, \; \incise, \; \DF)}
{\snd(\LS, \; \AS, \; \DFincised)}
{\{i+1; \; \K{\LS}{i+1}, \K{\AS}{i+1}, \K{\N}{i+1}\}}

\section{Mutations}
\label{sec:mutations}

When engaged in a surgical procedure (and, in general, in a security ceremony), humans might make mistakes because of various reasons, such as communication errors, distraction, inexperience, stress, etc. These mistakes alter the process flow and create deviations of the original ceremony specification that may impact the security of the ceremony or, in a surgical procedure, the safety of the patient. 

We adapt and extend to surgical procedures the approach of~\cite{SempreboniVigano2022}, which allows security analysts to model mistakes by human agents as \emph{mutations} of the behaviour that the ceremony originally specified for such agents.\footnote{These mutations refer to deviations from the expected sequence of actions in a process or procedure, not to the mutations found in other fields such as molecular biology or genetic mutations. We use the mutation names given in~\cite{SempreboniVigano2022} but other names have been proposed for similar mutations in different disciplines, e.g., in biology.} Mutations thus create alternative formal specifications of the original ceremony, which we can then formally analyse to see if they lead to violations of the intended properties (and thus endanger patient safety). Studying these mutations is also interesting as they might reveal alternative ways to carry out the procedure that do not endanger the patient and are, possibly, faster or more efficient. In this paper, we focus on formal analysis, dis-/proving properties of a procedure, but in the future we plan also to carry out a cost analysis of the different secure alternatives.

Since mutations allow humans to do things that were not foreseen in the original procedure, we formalise them by introducing new transition rules that are themselves mutations of the original ones.
For surgical procedures, we focus on two mutations, skip and replace, but our approach is open to other mutations, as we discuss in more detail below.

The \emph{skip mutation} allows us to formalise an agent skipping some actions that the surgical procedure expects them to carry out.
For instance, for the cutting stage, the case in which $\AS$ does not apply the clips but nonetheless 
sends a confirmation message signaling task completion is formalised by the mutated rule
\NewTransitions
{\{2; \; \; \K{\LS}{2}, \K{\AS}{2}, \K{\N}{2}\}}
{\rcv(\N, \; \AS, \; \QMX)}
{\ck(\QMX \; = \; \clipsprovided)}
{\stkout{\surgicalaction(\AS, \; \apply, \; \clips)}}
{\snd(\AS, \; \LS, \; \clipsapplied)}
{\{3; \; \; \K{\LS}{3}, \K{\AS}{3}, \K{\N}{3}\}}
and the case in which $\AS$ applies the clips but does not send a confirmation message is formalised by the mutated rule

\NewTransitions
{\{2; \; \; \K{\LS}{2}, \K{\AS}{2}, \K{\N}{2}\}}
{\rcv(\N, \; \AS, \; \QMX)}
{\ck(\QMX \; = \; \clipsprovided)}
{\surgicalaction(\AS, \; \apply, \; \clips)}
{\stkout{\snd(\AS, \; \LS, \; \clipsapplied)}}
{\{3; \; \; \K{\LS}{3}, \K{\AS}{3}, \K{\N}{3}\}}
\noindent

In the \emph{replace mutation}, an agent performs their actions as expected but replaces a message with another one. For instance, in case of complex messages consisting of different components, an agent could send just part of the message by mistake, as considered in the mutations in~\cite{SempreboniVigano2022}. For our case study, where messages are simple, we introduce the novel (w.r.t.~\cite{SempreboniVigano2022}) concept of \emph{negative message}, which we write as ``$\mathit{not}\_ m$'', e.g., $\notclipsapplied$, and we extend accordingly the set of messages agents can send or receive.
This allows agents in the cutting stage of our case study to execute mutated rules such as:
\NewTransitions
{\{2; \; \; \K{\LS}{2}, \K{\AS}{2}, \K{\N}{2}\}}
{\rcv(\N, \; \AS, \; \QMX)}
{\ck(\QMX \; = \; \clipsprovided)}
{\surgicalaction(\AS, \; \apply, \; \clips)}
{\snd(\AS, \; \LS, \; \notclipsapplied)}
{\{3; \; \; \K{\LS}{3}, \K{\AS}{3}, \K{\N}{3}\}}

Each action in a surgical procedure has a purpose and altering even a single action will cause some sort of propagation of the mistake for the next agents as well, which could impact patient safety. 
When a mutation happens, the other agents will not be able to carry out their actions unless their rules are mutated as well. For instance, if $\AS$ does not apply the clips, then $\LS$ will not cut and execution will deadlock. Our aim is for a procedure not to deadlock during execution but rather to be executed completely so that we can check whether the intended properties are satisfied even in presence of a mistake. To ensure that we only have executable traces, every mutation is matched via a \emph{matching mutation} and propagated through a trace. 

A {matching mutation for a skip mutation} depends on the ability of an agent to perform their action given that the previous agent has skipped theirs. For example, if $\N$ skips their action to provide the clips

\NewTransitions
{\{1; \; \; \K{\LS}{1}, \K{\AS}{1}, \K{\N}{1}\}}
{\rcv(\LS, \; \N, \; \mathit{\QMX})}
{\ck(\mathit{\QMX} \; = \; \clipsrequested)}
{\stkout{\surgicalaction(\N, \; \provide, \; \clips)}}
{\snd(\N, \; \AS, \; \clipsprovided)}
{\{2; \; \; \K{\LS}{2}, \K{\AS}{2}, \K{\N}{2}\}}
the matching mutation formalises that the action 
$\AS$ was going to perform, apply clips, is skipped as it is impossible for $\AS$ to apply clips unless $\N$ provides them 

\NewTransitions
{\{2; \; \; \K{\LS}{2}, \K{\AS}{2}, \K{\N}{2}\}}
{\rcv(\N, \; \AS, \; \mathit{\QMX})}
{\ck(\mathit{\QMX} \; = \; \clipsprovided)}
{\stkout{\surgicalaction(\AS, \; \apply, \; \clips)}}
{\snd(\AS, \; \LS, \; \clipsapplied)}
{\{3; \; \; \K{\LS}{3}, \K{\AS}{3}, \K{\N}{3}\}}
If an agent skips the sending of a message, the corresponding matching mutation ensures that the agent who was intended to receive that message skips the receipt and the corresponding check, e.g., if $\AS$ skips $\snd(\AS, \; \LS, \; \clipsapplied)$, then
\NewTransitions
{\{3; \; \; \K{\LS}{3}, \K{\AS}{3}, \K{\N}{3}\}}
{\stkout{\rcv(\AS, \; \LS, \; \mathit{\QMX})}}
{\stkout{\ck (\mathit{\QMX} \; = \; \clipsapplied)}}
{\surgicalaction(\LS, \; \cut, \; \ureter )}
{\snd(\LS, \; \AS, \; \cutdone)}
{\{4; \; \; \K{\LS}{4}, \K{\AS}{4}, \K{\N}{4}\}}
Note that this will have an effect on the knowledge of the agent, e.g., since $\LS$ does not receive the $\clipsapplied$ message, their knowledge will not be increased.

Rules matching a replace mutation enable receipt of a negated message, e.g. 
\NewTransitions
{\{3; \; \; \K{\LS}{3}, \K{\AS}{3}, \K{\N}{3}\}}
{\rcv(\AS, \; \LS, \; \mathit{\QMX})}
{\ck (\mathit{\QMX} \; = \; \notclipsapplied)}
{\surgicalaction(\LS, \; \cut, \; \ureter )}
{\snd(\LS, \; \AS, \; \cutdone)}
{\{4; \; \; \K{\LS}{4}, \K{\AS}{4}, \K{\N}{4}\}}

See \secref{sec:MatchingMutations} for more details on matching mutations and their propagation.
We will discuss other possibly interesting mutations in the conclusions.


\section{Formal and Automated Analysis}
\label{sec:formal-analysis}

\subsection{Properties}
\label{sec:properties}
Surgical procedures should first and foremost guarantee patient safety. Hence, everything that might endanger it should be avoided and formally specified as a \emph{property} to be satisfied.
This way, we can formally analyse it and either prove it to hold or, if not, produce a trace that shows the sequence of actions violating the property. In other words, we should aim to verify a property or to falsify it. 
Falsification means searching for an attack trace, which the analysis tool outputs when one is found. If no trace is found and the entire search space has not been fully explored, the tool continues searching.
Generally, verifying that a ceremony satisfies a property (and thus that no attack trace exists) is not always possible, as the problem typically generates an infinite search space. Unless distinctive techniques are used such as those implemented in security protocol analysis tools such as Tamarin~\cite{meier2013tamarin} or ProVerif~\cite{BlanchetCSFW01}, the analysis tool (as it happens for our use of UPPAAL) will time out or only terminate if the scenario to be analysed is made finite.

In our approach, as is often done in security protocol analysis, we use a linear temporal logic to formalise properties. This logic allows us to specify that if an event occurs now, then certain other events must have occurred in the past.
For instance, the patient should not bleed out due to a negligent incision, i.e., a surgeon should only execute a cut after clips have been applied. Specifically, we require that in all traces, if $\LS$ carries out a cut at some time instant, then there must exist previous time instants, ordered temporally, in which the clips have been requested, provided and applied:

\begin{property}[Clip-before-cutting]\label{property-one} 
For all traces,
\begin{align*}
    \surgicalaction(\LS, \cut) \; @ l   
    \ \Longrightarrow \ & \surgicalaction(\LS, \; \request, \; \clips) \; @ i \\
    & \& \ \, \surgicalaction(\N, \; \provide, \; \clips) \; @ j \\
    & \& \ \, \surgicalaction(\AS, \; \apply, \; \clips) \; @ k \\
    & \& \ \, i<j<k<l
\end{align*}
\end{property}

Property~\ref{property-one}, which is a general and quite obvious property of any surgical procedure but is also explicitly inspired by the informal discussion in~\cite{Guillonneau2000}, establishes a fundamental sequence of actions that occur in any procedure that includes a cutting stage (cf.~Fig.~\ref{fig:msc_a}).
The following three properties consider, instead, the lateral dissection stage (cf.~Fig.~\ref{fig:msc_b}) and are again inspired by~\cite{Guillonneau2000} as well as by the more recent~\cite{Moschovas2022-ff,Tzelves2022-gm}.

Property~\ref{property-two} pertains to the dissection of the pedicle. Due to the presence of arteries and veins that pose a significant risk of bleeding, it is imperative for the surgeon to exercise caution regarding the posterolateral neurovascular bundles during this dissection. 
We formalise this by requiring in all traces that if $\LS$ performs this dissection at some time instant, then there must be preceding instants at which the vas deferens and seminal vesicles have been retracted (for optimal exposure of the pedicle), and the pedicle has been identified and cauterised.

\begin{property}[Dissection of the pedicle]\label{property-two} 
For all traces,
\begin{align*}
    \surgicalaction(\LS, \dissect, \pedicle) \; @ l   
    \ \Longrightarrow \ & \surgicalaction(\LS, \; \pull, \; \VDandSV) \; @ i \\
    & \& \ \, \surgicalaction(\AS, \; \lookfor, \; \pedicle) \; @ j \\
    & \& \ \, \surgicalaction(\AS, \; \cauterise, \; \pedicle) \; @ k \\
    & \& \ \, i<j<k<l
\end{align*}
\end{property}

Property~\ref{property-three} checks whether the incision of DF has been performed. This incision must be extended to the point where the bundles enter the pelvic muscular floor, which is crucial for the subsequent stage of the surgery. We formalise this by requiring that in all traces where $\LS$ incises the DF at a time instant, there must be preceding  instants during which the pericapsular fatty space was entered, the visceral fascia was incised, and the NVB were inspected.

\begin{property}[Incision of the Denonvilliers' Fascia]\label{property-three} 
For all traces,
\begin{align*}
    \surgicalaction(\LS, \incise, \DF ) \; @ l   
    \ \Longrightarrow \ & \surgicalaction(\LS, \enter, \PFS) \; @ i \\
    & \& \ \, \surgicalaction(\LS, \incise, \visceralfascia) \; @ j \\
    & \& \ \, \surgicalaction(\AS, \inspect, \NVB) \; @ k \\
    & \& \ \, i<j<k<l
\end{align*}
\end{property}

Property~\ref{property-four} is essential for assessing patient outcomes, as preserving the NVB is crucial for potency recovery \cite{Moschovas2022-ff}. To ensure the NVB remain intact, 
we require
that in all traces where $\AS$ inspects the NVB at a given time instant, there must be preceding instants during which the pedicle has been cauterised, the pericapsular fatty space has been entered and the visceral fascia has been incised.

\begin{property}[Check if the nerves are preserved]\label{property-four} 
For all traces,
\begin{align*}
    \surgicalaction(\AS, \inspect, \NVB) \; @ l   
    \ \Longrightarrow \ & \surgicalaction(\AS, \; \cauterise, \; \pedicle) \; @ i \\
    & \& \ \, \surgicalaction(\LS, \; \enter, \; \PFS) \; @ j \\
    & \& \ \, \surgicalaction(\LS, \; \incise, \; \visceralfascia) \; @ k  \\
    & \& \ \, i<j<k<l
\end{align*}
\end{property}

The above are examples of the kinds of properties that we can prove for surgical procedures as they guarantee the safety of the patient. Other interesting properties could be considered, especially when focusing on the other stages of the procedure or on more complex procedures. For instance, the above properties are all trace-based properties, showing that if an event holds in a state of a trace, then other events hold in other states of the trace, but one could also consider state-based properties, i.e., properties requiring that something holds in a state of a trace (this is particularly relevant for secrecy of information when messages are encrypted, which we have omitted to do in this first discussion of our approach but which we will consider in future work, especially in the case of telesurgeries).

\subsection{Automated Analysis Using UPPAAL}
\label{sec:analysis-with-UPPAAL}

To automate the analysis of the intended properties, we have used UPPAAL, an integrated tool environment for modeling, validation and verification of real-time systems (\url{https://uppaal.org}).\footnote{We chose UPPAAL for its visual simplicity but we could also have used the X-Men Tool, the extension of the Tamarin prover to mutation-based analysis of security ceremonies given in~\cite{SempreboniVigano2022}.}
We lack space to give the UPPAAL encodings of the formal models of the agents, which encode the original transition rules and the mutated ones and thus require much space, but we give them in \secref{app:formal-analysis}.

Encoding a model of the full laparoscopic prostatectomy, including the two stages we considered here and the other stages of the procedure, is well beyond what UPPAAL's analysis engine can handle as it would need to search through too many states, so in future work we will investigate the definition of abstractions or optimisations to avoid state space explosion. In this paper, we have thus considered the two stages independently, providing two sets of encodings: for each stage, we encode the models of the corresponding agents (three for the cutting stage and two for the lateral dissection stage) as well as an encoding of an additional component that represents the entire flow of that stage, which UPPAAL requires to allow the agents to synchronously work together.

For each agent, we modelled the states they can reach, the preconditions to reach these states and what transitions they can take to reach these states, including the original transition rules and the mutated ones. All mutations and matching mutations are explicitly constructed and labeled for clarity and ease of understanding (the focus of our approach is to diverge from concealing the mistakes that might emerge because of the mutations and instead delineate pathways where property violations are likely to occur). The preconditions are called \emph{guards} and represent the knowledge an agent has to posses to take a transition. The expected sequence of transitions includes all the necessary guards. However, other possible transitions lack guards to ensure the model can operate under various scenarios involving one or multiple mutations. Including the necessary guards and modeling each transition for every possible mutation would result in redundant duplicates, unnecessarily extending the model.
Upon the execution of a transition, which signifies an agent has completed their surgical action, we update the variable associated with this action. We also update a message variable to convey this information, indicating that the agent is transmitting a message to inform the next agent that the action has been executed.

We used UPPAAL to analyse two kinds of properties: those that affirm the validity of a single trace and those that should be valid for all traces. Properties of the first kind serve as evidence of the model's executability (i.e., there is a trace that follows the expected sequential order).
We do not discuss them here as they are basic properties that are, in a sense, preconditions of more interesting analyses. 
Instead, we focus on properties intended to be valid across all traces so as to find out whether agent mistakes lead to vulnerabilities. 
If a property fails to hold, the tool furnishes us with a counterexample trace in its output.\footnote{Note that UPPAAL outputs only one trace that violates the property and cannot provide all possible ``attacks''.
Note also that mutations and matching mutations may yield violations that make little sense in real life, so UPPAAL outputs should be checked to see if the violation is a false positive or 
could indeed occur in real life.}

\begin{figure}[t]
    \centering
    \includegraphics[width=8.4cm]{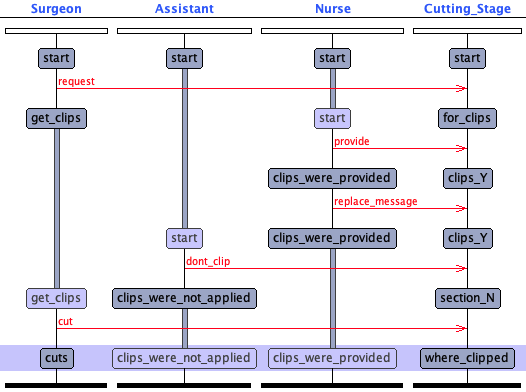}
    \caption{UPPAAL Output of the Analysis of Property~\ref{property-one} With Mutations Enabled}
    \label{fig:property_1_uppaal}
\end{figure}

Property~\ref{property-one} states that for all traces in which the surgeon performs a cut, there must be three previous time instants in which the clips have been requested, provided and applied. We input Property~\ref{property-one} in UPPAAL and the tool performed an automated analysis that confirmed that the property holds true across all traces when mutations are deactivated, but intriguing violations caused by errors made by agents become discernible upon enabling mutations (and matching mutations). These errors are represented as execution traces (output by UPPAAL) in which the property under analysis is not satisfied. For instance, UPPAAL outputs the ``attack'' trace depicted in Fig.~\ref{fig:property_1_uppaal}, which illustrates that the surgeon can execute a cut without the application of clips if mistakes occur due to miscommunication or negligence.

A similar result is observed for Property~\ref{property-two}, for which UPPAAL outputs an attack trace (shown in Fig.~\ref{fig:property_2_3_uppaal} in \secref{app:formal-analysis}) in which $\LS$ dissects the pedicle without vasa deferentes (VD) and seminal vesicles (SV) being pulled and without the pedicle being cauterised. This result could indicate several possibilities: the pedicle might have been visible without the need to pull VD and SV and there might have been no bleeding requiring cauterisation. Even though the patient's safety may not be compromised in this scenario, this trace still represents a violation of the expected sequence of actions. Therefore, it is crucial for a surgeon to interpret the results, as this trace could represent a legitimate shortcut in the procedure rather than an incorrect trace. Our approach thus allows clinicians to consider variants of the procedure and reason about them.

Property~\ref{property-three} is violated when $\LS$ makes an incision of the DF without incising the visceral fascia and without $\AS$ inspecting the NVB (the corresponding attack trace is shown in Fig.~\ref{fig:property_2_3_uppaal} in \secref{app:formal-analysis}).
If the NVB are not of interest, maybe the patient suffers from erectile dysfunction already, a surgeon might skip checking the state of the NVB and remove them if necessary. However, the trace output by UPPAAL would still be considered an attack trace 
as our model specifies that the NVB must be inspected before the final incision of the fascia is made.

For Property~\ref{property-four} UPPAAL outputs an attack (Fig.~\ref{fig:property_4_uppaal} in \secref{app:formal-analysis}) in which the NVB have been inspected without entering the pericapsular fatty space and without incising the visceral fascia. It is particularly important to ensure that Property~\ref{property-four} holds because of the presence of numerous NVB fibers between the posterior and intermediate layers of the DF, which makes dissection in this area highly hazardous for erectile nerves and should be avoided in nerve-sparing procedures~\cite{Tzelves2022-gm}.

\section{Concluding Remarks}
\label{sec:conclusions}

\emph{Summary.} 
We proposed a new approach to model and reason about, in a formal and automated way,  surgical procedures and the properties they should guarantee. Our approach relies on viewing surgical procedures as security ceremonies and leveraging well-established techniques developed for the analysis of such ceremonies. Specifically, we adapted and extended (e.g., with negative messages and using a different tool) the approach of~\cite{SempreboniVigano2022}. In doing so, we have provided not only formal models of the agents involved in a surgical procedure, but also formalised a way to reason about the possible mistakes the agents might make by viewing these mistakes as mutations of the original procedure. As concrete examples, we considered two stages of a laparoscopic prostatectomy. 
Our analysis has revealed a number of possible violations that can occur due to agent mistakes. 
While the identified violations are not particularly complex, they highlight the potential of our approach to uncover critical issues of surgical procedures.
Modelling all the steps of the whole surgical procedure, involving more agents, actions, messages, properties and possible mutations, will allow us to identify further violations but we postpone this to future work as the extended models quickly become too complex for UPPAAL to handle without the inclusion of abstractions or optimisations (which we are currently investigating). 

\emph{Future Work.} 
We view this paper as the first step towards the full-fledged analysis of surgical procedures. To that end, we plan to extend our approach in a number of ways. First, as just mentioned, to encompass a complete laparoscopic prostatectomy (and other procedures) by modelling and analysing all stages holistically rather than independently. We will explore various methodologies for performing a prostatectomy, involving an expanded set of agents, including both human and robotic agents, capable of executing a broader range of actions and transmitting additional messages. We will also consider telesurgeries, emphasising the importance of cryptography for secure communication between agents in the presence of possible attackers (and that is one of the main reasons why we adopted the security ceremony viewpoint). 

We also plan to consider other relevant mutations. For instance, it will be interesting to consider a mutation that ``negates'' an action by undoing it, which would, e.g., capture the mistake that occurs when the assistant applies the clips but then removes them before the surgeon cuts, resulting in bleeding. 

Our approach is open to all that and more. For instance, even though we could not discuss them in full detail here, our approach also allows one to explore variants of a procedure not only to identify potential attacks but also to study alternative approaches in which, say, the order of the actions is different from the ones in the minds of the surgical team. Our objective is to offer the most suitable surgical approach for each individual, tailoring the procedure to meet specific patient needs and conditions.

%
%



\appendix

\section{More Details on the Mutations of \secref{sec:mutations}}

In this section, we give mutated rules for the three agents of our case study, and discuss matching mutations and their propagation. The mutated rules can be carried out independently one by one or in conjunction. Moreover, rules of an agent are mutated to match a mutation caused by another agent, as well as to propagate that mutation through the execution.
Some of these combinations will result in trivial cases (e.g., if clips are not requested, then they will not be provided, they will not be applied, and no cut will be performed)
but other combinations might result in interesting violations of the intended properties, which our approach allows us to identify automatically.

\subsection{Skip and Replace Mutations} 
\label{app:SkipReplace}

The mutated rules that formalise the skip of a surgical action or of the sending of a message are obtained from the original transition rules simply by removing that action as shown in \secref{sec:mutations}.
To formalise the replace mutation, we first extend the sets $M_\Ag^s$ and $M_\Ag^r$ of messages an agent $\Ag$ can send and receive by including \emph{negative messages}, i.e., including $\mathit{not}\_ m$ for every $m$ in the set, where $\mathit{not}\_ \mathit{not}\_ m$ is simply $m$. 
In the replace mutation, agents perform their actions as expected but send the negation of the intended message as we have done in the example given in \secref{sec:mutations}.
As we remarked, this is one example of replace mutation that is not considered in~\cite{SempreboniVigano2022}, which instead 
focuses on mutations in which agents replace a composite message with a simpler one that includes only some of the components. We cannot consider such a mutation here as messages are solely constants, but it will be useful for when we extend to other stages of a surgical procedure.

\subsection{Matching Mutations}
\label{sec:MatchingMutations}

To match the mutation for skipping a surgical action, we need to consider two distinct cases. The first case occurs when agents can perform their actions even if the preceding agent has skipped theirs. The second case arises when an agent is unable to proceed unless we match the mutation that occurred previously.

Let us consider the cutting stage as an example; the lateral dissection stage is treated analogously.

\begin{description}
\item[Case 1:] even if the surgeon does not request clips, the nurse can still provide them since the surgeon's action is not a prerequisite for supplying clips.
\NewTransitionsShortend
{\{0; \; \; \K{\LS}{0}, \K{\AS}{0}, \K{\N}{0}\}}
{\stkout{\surgicalaction(\LS, \; \request, \; \clips)}} 
{\snd(\LS, \; \N, \; \clipsrequested)} 
{\{1; \; \; \K{\LS}{1}, \K{\AS}{1}, \K{\N}{1}\}}
\vspace*{-0.5cm}
\NewTransitions
{\{1; \; \; \K{\LS}{1}, \K{\AS}{1}, \K{\N}{1}\}}
{\rcv(\LS, \; \N, \; \QMX)}
{\ck(\QMX \; = \; \clipsrequested)}
{\surgicalaction(\N, \; \provide, \; \clips)}
{\snd(\N, \; \AS, \; \clipsprovided)}
{\{2; \; \; \K{\LS}{2}, \K{\AS}{2}, \K{\N}{2}\}}
\end{description}

\begin{description}
\item[Case 2:] if the nurse did not previously provide clips, the assistant would be unable to apply them. Therefore, the matching mutation would involve skipping the surgical action the assistant was meant to perform as well:
\NewTransitions
{\{1; \; \; \K{\LS}{1}, \K{\AS}{1}, \K{\N}{1}\}}
{\rcv(\LS, \; \N, \; \QMX)}
{\ck(\QMX \; = \; \clipsrequested)}
{\stkout{\surgicalaction(\N, \; \provide, \; \clips)}}
{\snd(\N, \; \AS, \; \clipsprovided)}
{\{2; \; \; \K{\LS}{2}, \K{\AS}{2}, \K{\N}{2}\}}
\NewTransitions
{\{2; \; \; \K{\LS}{2}, \K{\AS}{2}, \K{\N}{2}\}}
{\rcv(\N, \; \AS, \; \QMX)}
{\ck(\QMX \; = \; \clipsprovided)}
{\stkout{\surgicalaction(\AS, \; \apply, \; \clips)}}
{\snd(\AS, \; \LS, \; \clipsapplied)}
{\{3; \; \; \K{\LS}{3}, \K{\AS}{3}, \K{\N}{3}\}}
\end{description}

For the mutations in which the sending of a message is skipped or a message is replaced, we consider pairs of rules. The first rule denotes the mutation and the second rule the matching mutation, so that if the first agent changes their expected behaviour (not sending the message or sending a wrong one), the second agent is able to proceed with their actions. 
For instance, the mutations matching the skip of the sending of a message for the first pair ``surgeon and nurse'' (that includes the states from $\ST{}{0}$ to $\ST{}{2}$) are:

\NewTransitionsShortend
{\{0; \; \; \K{\LS}{0}, \K{\AS}{0}, \K{\N}{0}\}}
{\surgicalaction(\LS, \; \request, \; \clips)} 
{\stkout{\snd(\LS, \; \N, \; \clipsrequested)}} 
{\{1; \; \; \K{\LS}{1}, \K{\AS}{1}, \K{\N}{1}\}}

\NewTransitions
{\{1; \; \; \K{\LS}{1}, \K{\AS}{1}, \K{\N}{1}\}}
{\stkout{\rcv(\LS, \; \N, \; \QMX)}}
{\stkout{\ck(\QMX \; = \; \clipsrequested)}}
{\surgicalaction(\N, \; \provide, \; \clips)}
{\snd(\N, \; \AS, \; \clipsprovided)}
{\{2; \; \; \K{\LS}{2}, \K{\AS}{2}, \K{\N}{2}\}}
The other three pairs (``nurse and assistant'' for states $\ST{}{1}$ to $\ST{}{3}$, ``assistant and surgeon'' for states $\ST{}{2}$ to $\ST{}{4}$, and `` surgeon and assistant'' for states $\ST{}{3}$ to $\ST{}{5}$) are similar.

The mutations matching a replace mutation, enabling the receipt of a negated message, for the pair ``surgeon and nurse'' (states $\ST{}{0}$ to $\ST{}{2}$) are:

\NewTransitionsShortend
{\{0; \; \; \K{\LS}{0}, \K{\AS}{0}, \K{\N}{0}\}}
{\surgicalaction(\LS, \; \request, \; \clips)} 
{\snd(\LS, \; \N, \; \notclipsrequested)}
{\{1; \; \; \K{\LS}{1}, \K{\AS}{1}, \K{\N}{1}\}}

\NewTransitions
{\{1; \; \; \K{\LS}{1}, \K{\AS}{1}, \K{\N}{1}\}}
{\rcv(\LS, \; \N, \; \QMX)}
{\ck(\QMX \; = \; \notclipsrequested)}
{\surgicalaction(\N, \; \provide, \; \clips)}
{\snd(\N, \; \AS, \; \clipsprovided)}
{\{2; \; \; \K{\LS}{2}, \K{\AS}{2}, \K{\N}{2}\}}
Again, the other three pairs are similar.

\section{More details on the Formal Analysis of \secref{sec:formal-analysis}}
\label{app:formal-analysis}

\begin{figure}[!h]
    \centering
    \includegraphics[width=15cm]{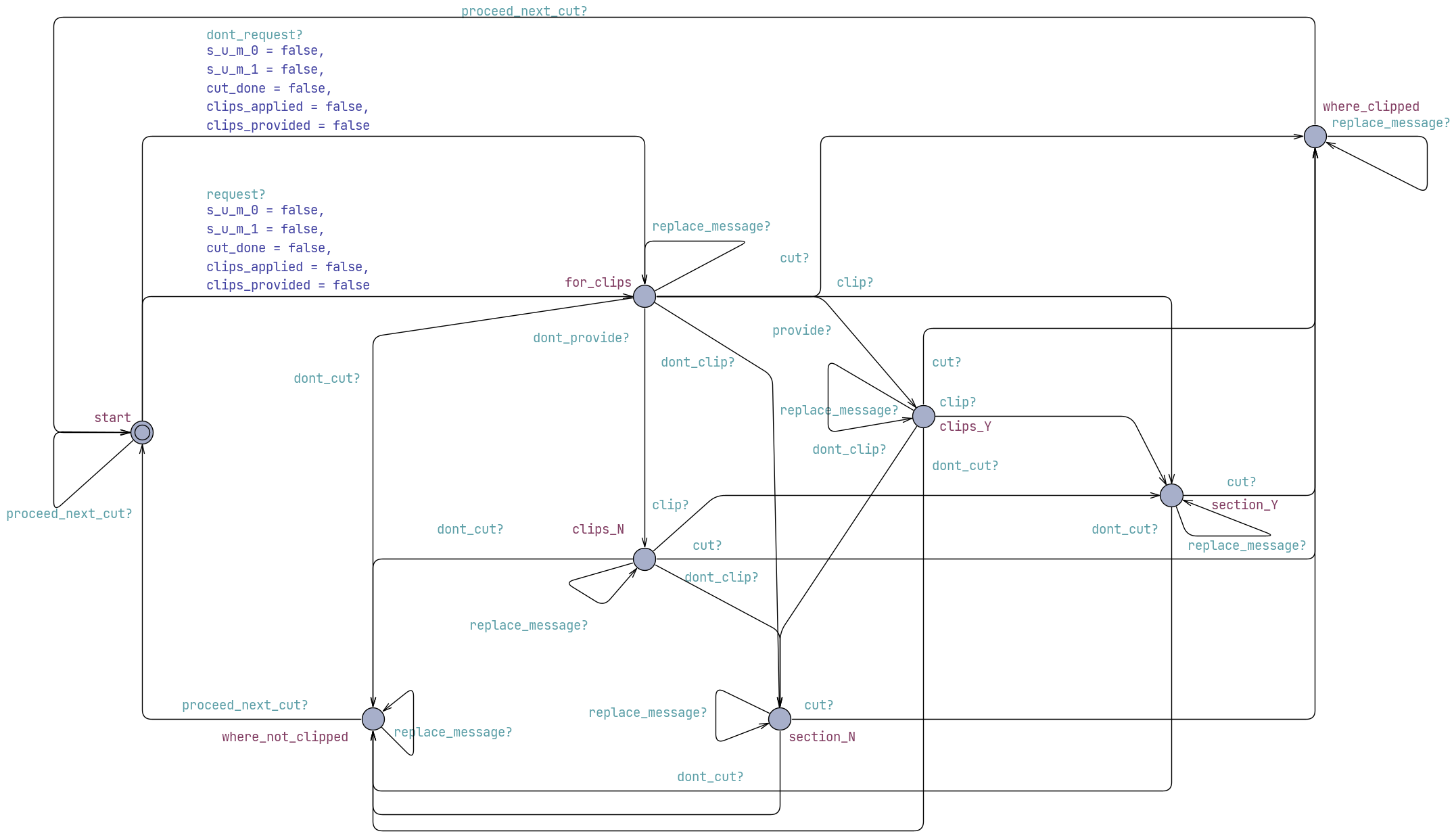}
    \caption{UPPAAL Encoding of the Model of the Cutting Stage}
    \label{fig:CS-uppaal-model}
\end{figure}

\begin{figure}[ht]
    \centering
    \includegraphics[width=15cm]{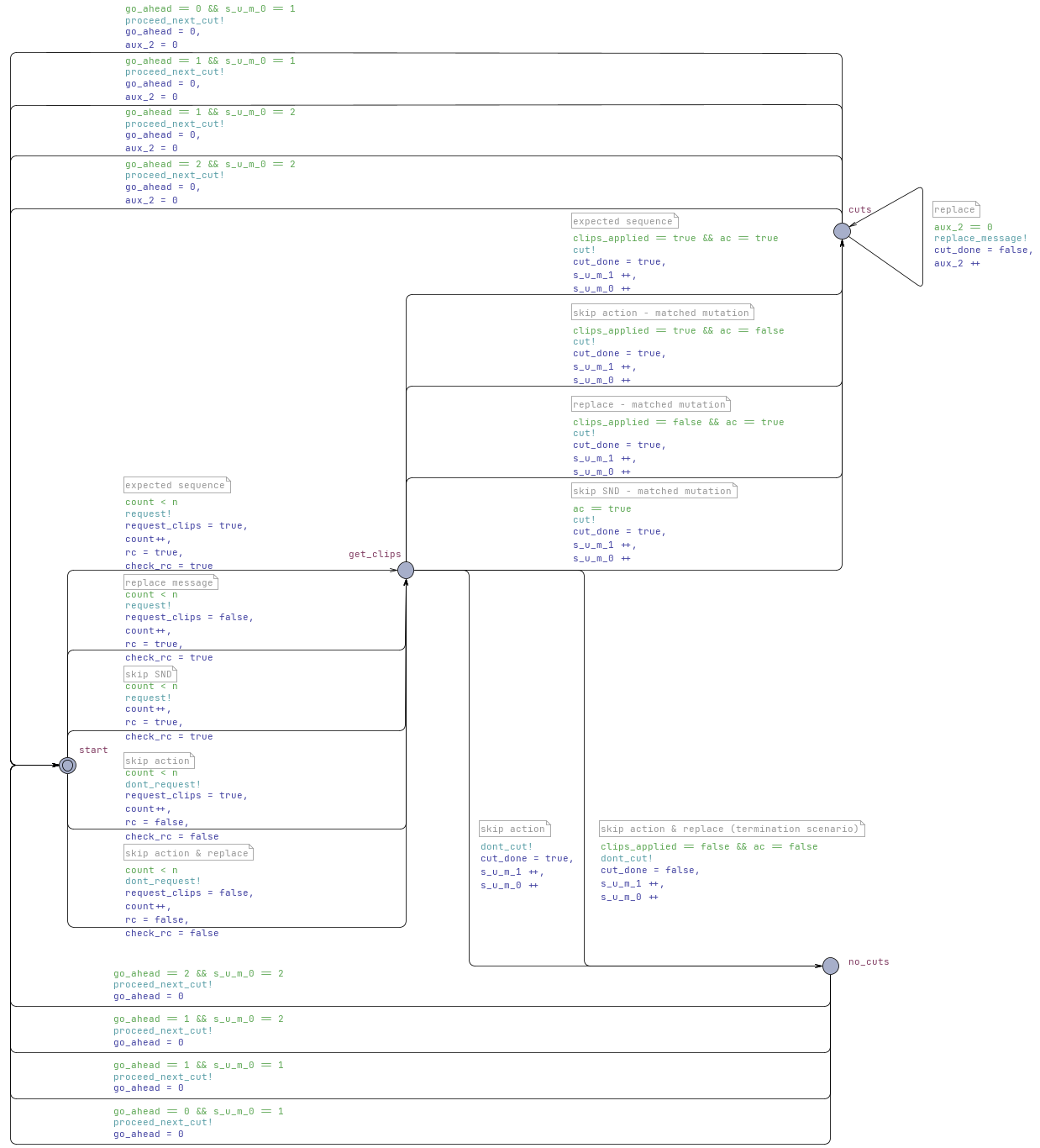}
    \caption{UPPAAL Encoding of the Model of the Surgeon for the Cutting Stage}
    \label{fig:LS-uppaal-model}
\end{figure}

\begin{sidewaysfigure}[h]
\vspace*{0cm}
    \includegraphics[height=4.5cm]{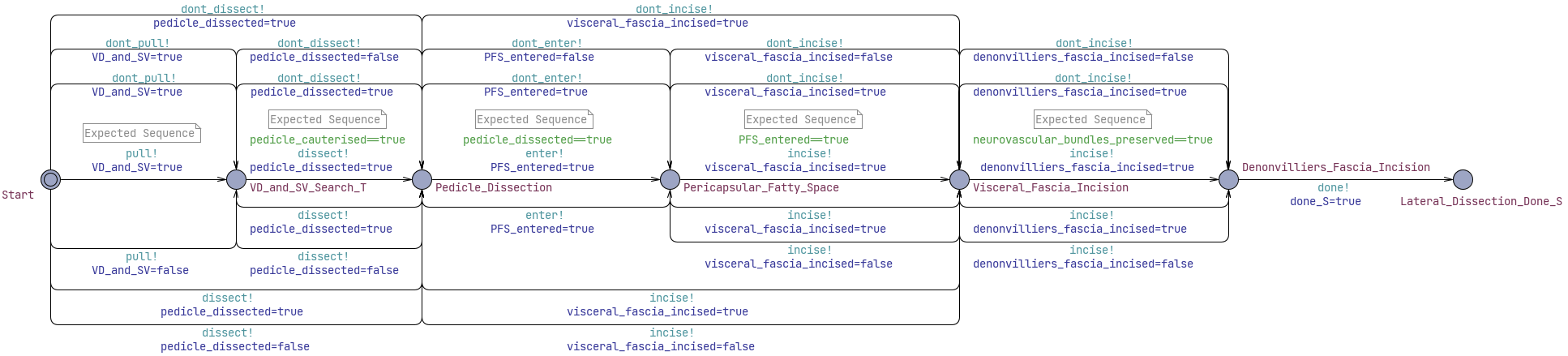} \\
    
    \includegraphics[height=4.5cm]{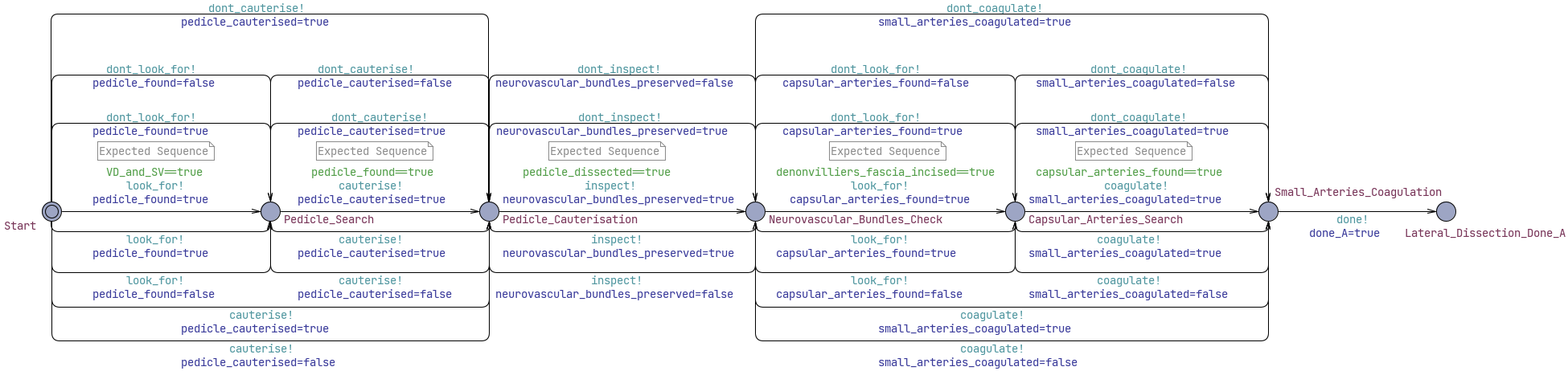} \\

    \includegraphics[height=2.4cm]{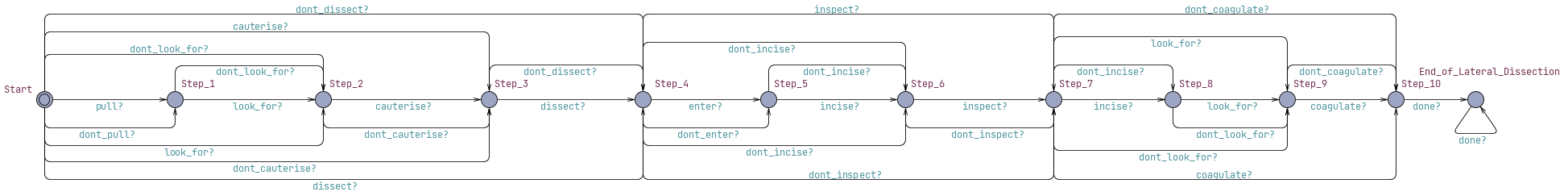}
    \caption{UPPAAL Encoding of the Models for the Lateral Dissection Stage}
    \label{fig:uppaal-model-dissection}
\end{sidewaysfigure}

In this appendix, we provide the UPPAAL encodings of the formal models of the agents.
Fig.~\ref{fig:CS-uppaal-model} shows the encoding of the overall model of the cutting stage, and Fig.~\ref{fig:LS-uppaal-model} shows the encoding of the model of the surgeon for the cutting stage.
Fig.~\ref{fig:uppaal-model-dissection} shows the encodings, for the lateral dissection stage, of the model of the surgeon, of the model of the assistant, and of the overall model.

\begin{figure}[ht]
\hspace*{-2cm}
    \includegraphics[width=8cm]{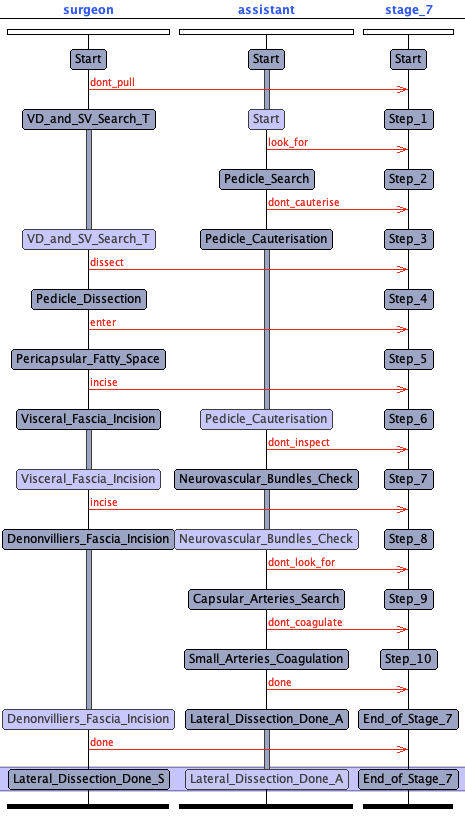}
    \includegraphics[width=8cm]{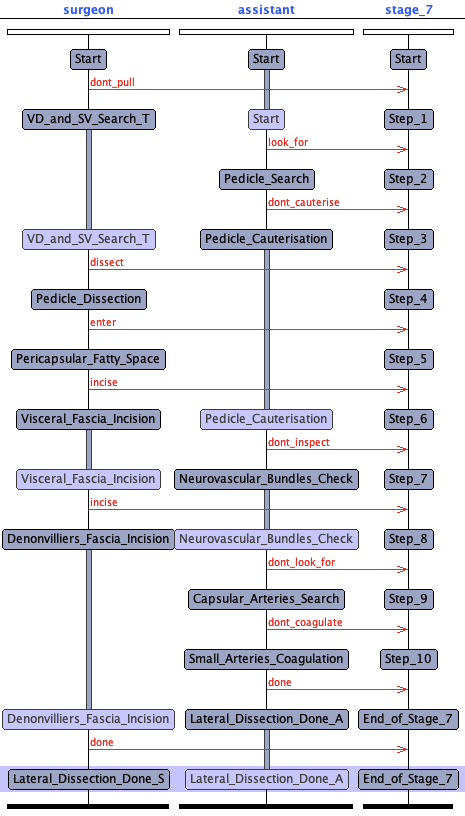}
    \caption{UPPAAL Output of the Analysis, With Mutations Enabled, of Property~\ref{property-two} and Property~\ref{property-three}}
    \label{fig:property_2_3_uppaal}
\end{figure}

\begin{figure}[ht]
    \centering
    \includegraphics[width=8.4cm]{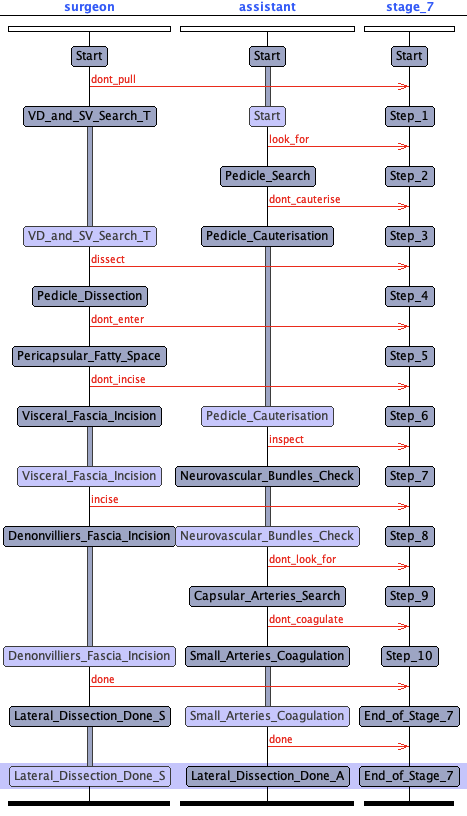}
    \caption{UPPAAL Output of the Analysis of Property~\ref{property-four} With Mutations Enabled}
    \label{fig:property_4_uppaal}
\end{figure}

Finally, 
Fig.~\ref{fig:property_2_3_uppaal} and 
Fig.~\ref{fig:property_4_uppaal} 
show the UPPAAL output of the analysis, with mutations enabled, of Property~\ref{property-two}, Property~\ref{property-three} and Property~\ref{property-four} .

\end{document}